\documentclass[lineno]{jfm}
\usepackage{graphicx}
\usepackage{bm}
\usepackage{epstopdf,epsfig}
\usepackage{newtxtext}
\usepackage{newtxmath}
\usepackage{natbib}
\usepackage{hyperref}
\usepackage{amsmath}
\usepackage{mathtools}
\usepackage{color}
\usepackage[normalem]{ulem}
\hypersetup{
    colorlinks = true,
    urlcolor   = blue,
    citecolor  = black,
}

\newcommand{\RomanNumeralCaps}[1]
\linenumbers

\title{Transition to the buoyancy dominated regime in a planar temporal forced
plume}

\author{L. Puggioni\aff{1,2}
  \corresp{\email{puggioni@lorentz.leidenuniv.nl}},
G. Boga\aff{3},
A. Cimarelli\aff{3},
M. Crialesi Esposito\aff{3},
S. Musacchio\aff{1},
E. Stalio\aff{3},
\and G. Boffetta\aff{1}}

\affiliation{\aff{1}Dipartimento di Fisica and INFN, Universit\`a di Torino, via Pietro Giuria 1, 10125 Torino, Italy
\aff{2}Instituut-Lorentz, Universiteit Leiden, P.O. Box 9506, 2300 RA Leiden, The Netherlands
\aff{3}Dipartimento di Ingegneria "Enzo Ferrari", Universit\'a di Modena e Reggio Emilia, via Vivarelli 10, 41125 Modena, Italy}

\begin{document}
\maketitle

\begin{abstract}
We study the transition from momentum- to buoyancy-dominated regime in 
temporal jets forced by gravity.
From the conservation of the thermal content and of the volume flux, 
we develop a simple model which is able to describe accurately 
the transition between the two regimes in terms of a single
parameter representing the entrainment coefficient.
Our analytical results are validated against a set of 
numerical simulations of temporal planar forced plumes at different 
initial values of Reynolds and Froude numbers. 
We find that, 
although the the pure jet-scaling law is not clearly observed in simulations
at finite Froude number, the model correctly describe the transition to 
the buoyancy-dominated regime which emerges at long times.
\end{abstract}


{\bf MSC Codes }  {\it(Optional)} Please enter your MSC Codes here

\section{Introduction}
\label{sec:headings}

Turbulent plumes are at the core of  several natural phenomena. For example,
plumes are formed during volcanic eruptions, and their dynamics controls the
transport of fine ash in the atmosphere \citep{woods2010turbulent}. In deep
oceans, water heats up during the formation of ocean crust, forming plumes
which can rise up to 300 m above the sea bed \citep{speer1989model}. Plumes
are also of central relevance in atmospheric cumulus convection
\citep{de2013entrainment}, which plays a fundamental role in the weather and
climate dynamic. The dynamics of plumes has been studied since the seminal,
analytical work of \cite{zeldovich1937asymptotic}, with an increasing
scientific interest in the last decades, producing a vast literature thoroughly
reviewed in \cite{hunt2011classical}. 

Plumes are usually described as "\textit{lazy}", if their motion is mostly
triggered by buoyancy and the initial momentum at their source is negligible, or
otherwise as "\textit{forced}" (or buoyant jets) if their momentum cannot be
neglected \citep{hunt2011classical,van2015energy}. Forced plumes become jets
when buoyancy effects are negligible.  By dimensional analysis, one obtains
that the ratio between forced and natural convection can be represented either
through the  Froude number  $Fr=U/\sqrt{g H}$, or the Richardson number
$Ri=Fr^{-2}$ , where $U$ is the characteristic flow velocity, $g$ is the
gravitational acceleration and $H$ is the characteristic flow scale (the
precise definition of these quantities will be provided in Section 
\ref{sec2}). Due to
the relevance of entrainment quantification in these flows, also the
flux-balance parameter \citep{morton1959forced}  is often used to
characterize the plume as lazy or forced, as it directly depends on the
entrainment coefficient (see \cite{van2015energy} for an exhaustive review on
the subject).  

Considerable effort has been devoted to the investigation of the local
phenomena driving entrainment. The exchange of momentum and scalar
concentration takes place across the region known as Turbulent/Non-Turbulent
Interface (TNTI), first examined by \cite{Corrsin_TNTI}, whose topology
determines the mixing and the growth of the turbulent region (see
\cite{daSilvaAnnuRev2014} for a detailed review). In this regard, the topology
of the TNTI has been thoroughly studied and characterized in jets
\citep{Westerweel_2005,daSilva_2010}, wakes \citep{Bisset_2002}, turbulent
boundary layers \citep{Borrell_Jimenez_2016,Watanabe2018}, shear layers
\citep{Watanabe_shearlayer} and other flow configurations. On the other hand,
several aspects concerning the dynamics of entrainment (e.g. nibbling and
engulfment processes) are still a subject of debate
\citep{Mathew2002,Westernweel2009, cimarelli2021numerical}. 
In \cite{krug2017global}, the authors reported that the 
increased rate of entrainment
observed in buoyant plumes compared with pure jets, 
may be due to the additional contribution
of the baroclinic torque and can be quantified to be in the order of ~10-15\%. 

At late times temporal plumes will always become lazy, but it is still unclear
how the transition from the forced to the  lazy regime occurs. 
In particular, it remains to be assessed whether the two regimes can be independently observed
and how the spreading evolve in the transition region.
Such transition
is difficult to describe in full-scale phenomena, as the spatially evolving
flow adds an intrinsic complexity. For example, a characterization was
performed by \cite{mazzino2021unraveling} where the authors studied the
turbulence properties of a puff (temporal jet propagating normally to buoyant
forces).
Another transition from momentum-dominated to buoyancy-dominated regime 
has been studied within the combination of Kelvin-Helmholtz and 
Rayleigh-Taylor instability and mixing 
\citep{snider1994rayleigh,akula2017dynamics} and recently investigated
in the turbulent regime 
\citep{brizzolara2021transition,brizzolara2023entrainment}.

A first level of simplification may be achieved considering the planar
configuration. In this setup, azimuthal symmetries are recovered in the
spanwise direction. In \cite{paillat2014entrainment}, the authors study the
entrainment in planar forced plumes, developing a theoretical model to describe
the effect of the planar shape on entrainment. Spatial planar plumes were also
mathematically characterized in \cite{van2014two1,van2014two}, where the
authors used the  scale-diagrams (originally presented in
\cite{morton1973scale}) to provide solutions for several types of plumes (both
lazy and forced) and jets. 

The complexity due to anisotropy and inhomogeneity can be further reduced by
adopting a temporally evolving setup, better suited to study the transition from the
momentum to the buoyancy-dominated regime. In this context, the planar and
temporally evolving configuration allows the use of periodic boundary
conditions and the recovery of the statistical homogeneity both in the
streamwise and in the spanwise directions. Despite its simplification, the
framework provided by temporally evolving flows allows to grasp the main flow
features. As shown in \cite{krug2017global} for a temporally evolving plume,
this flow paradigm is comparable to the corresponding spatially evolving flow
in the presence of a strong co-flow. The combination of planar and temporal
configuration has been used both in pure jets
(\cite{da2008invariants,vanReeuwijk_Holzner_2014}, 
\cite{cimarelli2021spatially}) and in plumes \citep{krug2017global}.

The present work addresses the properties of global scalings observed during
the transition from forced to lazy plumes, \emph{i.e.} from momentum to
buoyancy-dominated regimes. The topological and local properties of the
entrainment process, thoroughly studied in literature, are not considered here. 
In particular, the main focus is devoted to the characterization of the
transition regime itself. To this extent, by considering an initial flow
condition with a pre-imposed velocity profile of width $H_0$, it is know that
the time evolution of the width scales as $H(t)\propto t^{1/2}$ for the
momentum-dominated regime, while $H(t)\propto t$ is expected for the
buoyant-dominated regime. Whether these regimes can be clearly distinguished,
and if the transition occurs neatly or smoothly is the object of the present
work.

The remainder of this paper is organized as follows.
In Section~\ref{sec2} we present the mathematical model and the theoretical
predictions together with a summary of the numerical simulations.
In Section~\ref{sec3} we discuss the numerical results, while
Section~\ref{sec4} is devoted to the conclusions.
 
\section{Theoretical description and numerical methods}
\label{sec2}

\subsection{Flow settings and theoretical scaling}
\label{sec2.1}

We consider a planar, temporal forced plume with an initial velocity and temperature fields given by $u_x=u_y=0$ and by
\begin{equation}
u_z(x,y,z,t=0)= U_0 f(x/H_0)
\label{eq2.1}
\end{equation}
\begin{equation}
\theta(x,y,z,t=0)= T_0 f(x/H_0)
\label{eq2.2}
\end{equation}
where
  $x$ is the cross-stream direction,
  $y$ is the spanwise direction,
  $z$ is the streamwise direction, 
  $H_0$ is the initial width of the forced plume,
  $U_0$ is the initial velocity of the forced plume,
  $T_0$ is the difference between the temperature inside and outside the plume
  (which is assumed to be positive),
  and $f(\xi)$ (with $\xi=x/H_0$)
  is a symmetric, dimensionless function
  which has a limited support around
  $\xi=0$ such that $\int f(\xi) d\xi = 1$. 
By considering an incompressible flow and the Boussinesq approximation, the evolution of the velocity and the temperature fields is governed by
\begin{equation}
{\partial {\bm u} \over \partial t} + {\bm u} \cdot {\bm \nabla \bm u} +
{1 \over \rho_0} {\bm \nabla} p - \nu \nabla^2 {\bm u} + \beta {\bm g} \theta =0
\label{eq2.3}
\end{equation}
\begin{equation}
{\partial \theta \over \partial t} + {\bm u} \cdot {\bm \nabla} \theta 
- \kappa \nabla^2 \theta = 0
\label{eq2.4}
\end{equation}
where $\nu$ is the kinematic viscosity, $\kappa$ the thermal diffusivity, $\beta$ the thermal expansion coefficient and ${\bm g}=(0,0,-g)$. The dimensionless 
parameters which define the initial state are the Reynolds number 
$Re_0=U_0 H_0/\nu$, the Froude number $Fr_0=U_0/\sqrt{A_0 g H_0}$ and the
Prandtl number $Pr = \nu / \kappa$, where $A_0=\beta T_0/2$ is the Atwood 
number.  In the present work, all the simulations are done at $Pr =1$.

The system of equations (\ref{eq2.3}) and (\ref{eq2.4}) is solved in a cubic box of size $L$ by using periodic boundary conditions and the initial conditions (\ref{eq2.1}) and (\ref{eq2.2}). As a consequence, the flow is statistically homogeneous in the spanwise $(y)$ and streamwise $(z)$  directions, is statistically symmetric about the $x=0$ plane and the average operator (hereafter denoted as $\overline{(\cdot)}$) can be composed by a spatial average in the $y$-$z$ plane and by an ensemble average over independent samples. In accordance with such statistical symmetries of the flow, the equations for the mean flow solution read
\begin{equation}
{\partial \overline{u_z} \over \partial t} =
\nu {\partial^2 \overline{u_z} \over \partial x^2}
-{\partial \overline{u_z u_x} \over \partial x}
+\beta g \overline{\theta} \, ,
\label{eq2.7}
\end{equation}
\begin{equation}
{\partial \overline{\theta} \over \partial t}=
\kappa {\partial^2 \overline{\theta} \over \partial x^2}
-{\partial \overline{\theta u_x} \over \partial x} \, .
\label{eq2.8}
\end{equation}

Integration over $x$ gives the equations for the total
thermal content $C(t) \equiv \int_L \overline{\theta}(x,t) dx\equiv C_0$
and the volume flux $Q(t) \equiv \int_L \overline{u_z}(x,t) dx$
which obey to
\begin{equation}
{d Q \over d t} = \beta g C
\label{eq2.9}
\end{equation}
\begin{equation}
{d C \over d t} = 0
\label{eq2.10}
\end{equation}
with the obvious solutions
\begin{equation}\label{eq2.11}
Q(t)=Q_0 + \beta g C_0 t
\end{equation}
\begin{equation}\label{eq2.12}
C(t) = C_0 \, .
\end{equation}
Given the expressions (\ref{eq2.1}-\ref{eq2.2}) for the initial
profiles we have an explicit expression for the integration constants
$C_0=T_0 H_0$ and $Q_0=U_0 H_0$.
Equation (\ref{eq2.11}) indicates the presence of two different
regimes in time. Initially the constant term $Q_0$ dominates the flow rate: this is 
the shear (or jet) regime.
At later times buoyancy increases the total momentum together with the flow rate $Q$:
this is the buoyancy (or plume) regime.
The transition between the two regimes is expected to occur
when the two terms in (\ref{eq2.11}) are of the same order,
i.e. at the crossover time 
\begin{equation}
  t_g = \frac{U_0}{\beta g T_0}= \frac{Fr_0^2}{2} \, t_0
  \label{eq_buoyancy_time}
\end{equation}
where $t_0=H_0/U_0$ is the shear time. The above relation shows that the
transition time to the buoyancy regime increases quadratically with the Froude number.
We remark that an equivalent crossover time has been introduced in the 
case of unstably stratified shear layers in \cite{brizzolara2021transition}.

In analogy with the initial condition, we rewrite the total thermal content and
volume flux as follow:
\begin{equation}
Q(t)=U(t)\, H_U(t)
\label{eq_Hu}
\end{equation}
\begin{equation}
C_0=T(t)\, H_{T}(t)
\label{eq_Ht}
\end{equation}
The characteristic velocity $U(t)$ and temperature $T(t)$ are defined as:
\begin{equation}
  U(t) = \frac{1}{H_U(t)} \int_L \overline{u_z}(x,t) dx
  \label{eq_Ut}
\end{equation}
\begin{equation}
  T(t) = \frac{1}{H_T(t)} \int_L \overline{\theta}(x,t) dx\;.
  \label{eq_Tt}
\end{equation}
The length scales $H_{U}(t)$ and $H_{T}(t)$
represent the width at time $t$ of the velocity and temperature profiles, respectively. 
Their precise definition is arbitrary, provided that is consistent
with the initial value $H_{U}(t=0)=H_{T}(t=0)=H_0$.
In Sect.\ref{sec3.1} we will adopt a definition based on the
full-width-half-maximum of the profiles.  
By inserting (\ref{eq_Hu}) into (\ref{eq2.11}) we obtain
\begin{equation} 
U(t)\, H_U(t) = U_0 H_0 + 2 g A_0 H_0 t
  \label{eq_UHt}
\end{equation}
which relates the time evolution of $H_U(t)$ and $U(t)$.
Introducing the time-dependent Reynolds number $Re(t) = U(t)H_U(t)/\nu$,
the exact relation (\ref{eq_UHt}) can be rewritten as: 
\begin{equation}
Re(t)=Re_0 \left (1 + \frac{t}{t_g} \right )
  \label{eq_Ret}
\end{equation}
In order to derive an explicit prediction for the growth of $H_U(t)$
we further assume the following dimensional relation:   
\begin{equation}
  \frac{dH_U(t)}{dt}=\alpha U(t)
  \label{eq_closure}
\end{equation}
 where $\alpha$ represents the entrainment coefficient of the process.
The physical motivation for this simple dimensional relation is that the
growth of the width of the plume is due to the entrainment process, which is
driven by the turbulent flow inside the plume.  The hypothesis of
self-similarity of the flow implies that  the typical intensity of the
turbulent velocity fluctuations is proportional to the mean velocity $U(t)$.  
It is worth to notice that, in principles, the entrainment coefficients
in the momentum-dominated and buoyancy-dominated regimes could be different.
As an example, two distinct values of the entrainment coefficients
have been observed in unstable stratified shear layers, depending on the regime
\citep{brizzolara2023entrainment}.
Nonetheless, the assumption of a single value for the entrainment coefficient in our model
has the advantage to allow to derive explicit predictions for the time evolution of $H(t)$ and $U(t)$.    

Combining the relations~(\ref{eq_UHt}) and (\ref{eq_closure})
we obtain an explicit expression for the growth of the mixing layer
\begin{equation}
H_U(t) = H_0 \sqrt{1 + 2 \alpha \, \left (\frac{t}{t_0} \right ) + \frac{2\alpha}{Fr_0^2} \, \left (\frac{t}{t_0}\right )^2}
\label{eq:plume_width_scaling_t0}
\end{equation}
that can be rearranged to highlight the dependence on the crossover time
\begin{equation}
H_U(t) = H_0 \sqrt{1 + \alpha \, Fr_0^2 \left (\frac{t}{t_g}\right) + \frac{\alpha}{2}\, Fr_0^2 \left (\frac{t}{t_g}\right )^2}
\label{eq:plume_width_scaling}
\end{equation}
Inserting Eq.~(\ref{eq:plume_width_scaling}) in the exact relation
(\ref{eq_UHt}) we obtain a prediction for $U(t)$:
\begin{equation}
  U(t)=U_0 \frac{H_0}{H_U(t)}\left (1 + \frac{t}{t_g} \right )
  \label{eq:plume_velocity_scaling}
\end{equation}
Moreover, inserting (\ref{eq_Ht}) into (\ref{eq2.12}) we get $T(t) = T_0 H_0 /H_T(t)$.
  Assuming the the width of the temperature profile $H_T(t)$ is proportional to $H_U(t)$,
  so that the ratio $\gamma = H_U(t)/H_T(t)$ is constant,
  we can exploit the model~(\ref{eq:plume_width_scaling})
  to obtain scaling predictions for $H_T(t)$ and $T(t)$. 

From Eq.~(\ref{eq_buoyancy_time}) we get that, if $Fr_0 \gg 1$,
the crossover time $t_g$ is much longer than the shear time $t_0$.
In this case, from the expressions (\ref{eq:plume_width_scaling_t0}) and (\ref{eq:plume_width_scaling})
it is possible to identify two different scaling regimes that follow one another over time.
In the range of times $t_0 \ll t \ll t_g$,
the constant term $1$ and the quadratic term $(\alpha/2) Fr_0^2 (t/t_g)^2$
in (\ref{eq:plume_width_scaling}) can be neglected,
and the evolution is dominated by the initial momentum. 
This leads to the following temporal scaling laws
\begin{eqnarray}\label{eq:jetscaling}
H_U(t) = \gamma H_T(t) & \simeq & \sqrt{2\alpha} H_0 \left({t/t_0} \right)^{1/2} \nonumber \\
U(t) & \simeq & \frac{U_0}{\sqrt{2\alpha}} \left({t/t_0} \right)^{-1/2} \\
T(t) & \simeq & \gamma \frac{T_0}{\sqrt{2\alpha}} \left({t/t_0} \right)^{-1/2} \nonumber
\end{eqnarray}
The scaling law for $T(t)$ is a consequence of the constancy of $C(t)=C_0$.

These scaling laws are clearly transient, because the
quadratic term $(\alpha/2) Fr_0^2 (t/t_g)^2$
in (\ref{eq:plume_width_scaling})
eventually becomes dominant for $t \gg t_g$. 
Therefore, at long times we expect that the system recovers the plume regime
\begin{eqnarray}\label{eq:plumescaling}
H_U(t) = \gamma H_T(t) & \simeq & \sqrt{\frac{\alpha}{2}} H_g \left({t/t_g} \right) \nonumber \\
U(t) & \simeq & \sqrt{\frac{2}{\alpha}} U_g \\
T(t) & \simeq & \gamma T_g \sqrt{\frac{2}{\alpha}} \left({t/t_g} \right)^{-1} \nonumber
\end{eqnarray}
where $H_g=H_0(U_0/U_g)=H_0 Fr_0$, $U_g=(A_0 g H_0)^{1/2}$ and $T_g=T_0(U_g/U_0) = T_0/Fr_0 $
also known as free-fall quantities.
We will refer to equations (\ref{eq:jetscaling}) and (\ref{eq:plumescaling}) as
the jet and the plume temporal scaling respectively.

  We note that the model (\ref{eq:plume_width_scaling}) allows to define exactly the time of the transition between these regimes
  by intersecting the asymptotic laws for $H_U(t)$.
  The transition is expected to occur at $t=2t_g$,
  the mixing thickness and the velocity at the transition is 
  \begin{equation}
  	\begin{array}{r@{}l}
		H_U(2t_g) & {}= \sqrt{1+ 4\alpha} H_0~, \\ [5pt]
		U(2t_g)   & {}= (3/ \sqrt{1+ 4\alpha}) U_0 .
  	\end{array}
  \label{eq:scalings_trans}
  \end{equation}

  Beside the prediction for the global quantities of the system $H_{U,T}$, $U$ and $T$,
  the assumption of self-similarity of the turbulent flow allows to derive also     
  dimensional scaling predictions for other statistical quantities,
  such as the kinetic energy dissipation rate $\varepsilon_\nu = \nu \langle (\partial_j u_i)^2 \rangle$
  and the RMS turbulent velocity fluctuations  $u' = \sqrt{2 k/3 }$ where 
  $k = \langle \frac{1}{2} |{\bm u}-\overline{\bm u}|^2 \rangle $ is the turbulent kinetic energy 
  (here and in the following $\langle \cdot \rangle$ indicates the spatial average
  in the region $|x| \le H_U(t)$). 
  From dimensional analysis one has $\varepsilon_\nu(t) \simeq U(t)^3/H_U(t)$,
  while the scaling of velocity fluctuations $u'(t)$ is assumed to be proportional to that of $U(t)$. 
  Combining the scaling for $\varepsilon_\nu$ and $u'$ we can also derive a prediction
  for the Taylor-microscale Reynolds number $Re_\lambda = u' \lambda/\nu = u'^2\sqrt{15/ \nu \varepsilon_\nu}$,
  where $\lambda = \sqrt{15\nu/\varepsilon_\nu} u'$ is the Taylor microscale. 
  In the momentum dominated regime $t_0 \ll t \ll t_g$ we obtain
\begin{eqnarray}\label{eq:jetscaling_turb}
\varepsilon_\nu(t) &\simeq & \frac{U_0^3}{H_0} \left({t/t_0} \right)^{-2} \nonumber \\
Re_\lambda & \simeq  & \sqrt{Re_0} 
\end{eqnarray}
and in the buoyancy-dominated regime $t \gg t_g$ we get
    \begin{eqnarray}\label{eq:plumescaling_turb}
\varepsilon_\nu(t) &\simeq & \frac{U_0^3}{H_0 Fr_0^4} \left({t/t_g} \right)^{-1} = \frac{U_g^2}{t} \nonumber \\
Re_\lambda & \simeq  & \sqrt{Re_0} \left({t/t_g} \right)^{1/2} 
  \end{eqnarray}  

We remark that, formally, the pure jet scaling laws (\ref{eq:jetscaling})
can be observed only in the case of non-buoyant temporal jet ($Fr_0=\infty$).
Therefore, if the initial Froude number is $Fr_0 \le 1$ the jet scaling cannot be observed and the scaling (\ref{eq:plumescaling}) holds at all times. 
Moreover, at finite $Fr_0$, we expect that the jet term
$\alpha Fr_0^2(t/t_g)$  
in (\ref{eq:plume_width_scaling})
is contaminated by both the constant term $1$  
at short times $t \le t_0$
and by the buoyancy term $(\alpha/2)Fr_0^2(t/t_g)^2$ 
for $t \ge t_g$, thus preventing the observation of clear jet 
scaling regime (\ref{eq:jetscaling}), 
unless there is a large scale separation between $t_g$ and $t_0$, i.e., $Fr_0 \gg 1$.

\subsection{Direct Numerical Simulations}
\label{sec2.2}
Direct numerical simulations of equations (\ref{eq2.3}-\ref{eq2.4})
have been carried out in
a cubic periodic domain of size $L^3=\left( 2\pi \right)^3$ with $N^3=512^3$
equally spaced gridpoints, by means of a fully-parallel standard
pseudo-spectral code with a 2/3 dealiasing rule. Time integration is performed
using a second order Runge-Kutta scheme, with implicit integration of 
linear dissipative terms.
The initial conditions are given by (\ref{eq2.1}) and (\ref{eq2.2}) with the initial width of the forced plume fixed at $H_0=L/32$ and
\begin{equation}
f(\xi) = {1 \over 2} \left[ 1 + \tanh\left( {1 - 2 |\xi| \over 2 \sigma} \right) \right]
\label{eq2.16}
\end{equation}
where $\xi = x/H_0$ and $\sigma=5/56$. A uniform random noise with relative 
amplitude $0.04$ is superimposed to the velocity and temperature fields
into the inner region, in order to trigger the initial instability.

\begin{table}
\begin{center}
\begin{tabular}{lcc|ccccc}
$Set$ & $Fr_0$ & $Re_0$ & $A_0g$ & $U_0$ & $U_g$ & $t_0$ & $t_g$ \\[3pt]
A & 3.6      & 393  & 0.10 & 0.5 & 0.14 & 0.4 &  5 \\
B & 5.1      & 785  & 0.20 & 1.0 & 0.20 & 0.2 &  5 \\
C & 7.1      & 785  & 0.10 & 1.0 & 0.14 & 0.2 & 10 \\
D & 10.1     & 785  & 0.05 & 1.0 & 0.10 & 0.2 & 20 \\
E & 14.3     & 1571 & 0.10 & 2.0 & 0.14 & 0.1 & 20 \\
F & 20.2     & 1571 & 0.05 & 2.0 & 0.10 & 0.1 & 40 \\
O & $\infty$ & 758  & 0.00 & 1.0 & 0.00 & 0.2 & $\infty$ \\
\end{tabular}
\caption{
  Parameters of the different sets of simulations, ordered with increasing $Fr_0$. 
  Beside the nondimensional parameters $Fr_0 = U_0/U_g$ and $Re_0 = U_0 H_0/\nu$,
  we report the numerical values (in arbitrary units) of the dimensional parameters of the simulations:
  $A_0g$, $U_0$, $U_g = \sqrt{A_0 g H_0}$, $t_0 = H_0/U_0$, $t_g=U_0/A_0 g$. 
  The values of the other parameters are common to all simulations:
  $N= 512$, $L=2 \pi$, $H_0=L/32$, $T_0 = 1$, $\nu=\kappa = 2.5 \times 10^{-4}$.
}
\label{table1}
\end{center}
\end{table}

Six different sets of simulations have been carried out, each one with 10 different realizations of the initial random perturbation to compute statistics on appropriate sample sizes. 
Each set is characterized by different values of $U_0$, 
corresponding to different $Re_0$ and $Fr_0$, while $T_0$, $H_0$ and 
$\nu = \kappa$ are kept fixed.
An additional set of simulations with $A_0 g=0$, corresponding to the 
case without buoyancy (\textit{i.e.} a temporal jet with the temperature 
as a passive scalar) has been carried out for comparison. 
The complete set of parameters is listed
in table \ref{table1}.
The characteristic scales $H_U$ and $H_T$ are computed as 
the full-width-half-maximum lengths, i.e. 
$\overline{u_z}(H_U,t) = \overline{u_z}(0,t)/2$ and 
$\overline{\theta}(H_T,t) = \overline{\theta}(0,t)/2$.
This definition is consistent with the initial values $H_U(t=0)=H_T(t=0)=H_0$,
and it has the advantage that the value of $H_U$ and $H_T$ remains
constant (and equal to $H_0$) during the initial diffusive phase of the evolution.
All the characteristic quantities reported
and discussed are averaged over the 10 realizations of each set. 

\begin{figure}
\centering
\includegraphics[width=1\linewidth]{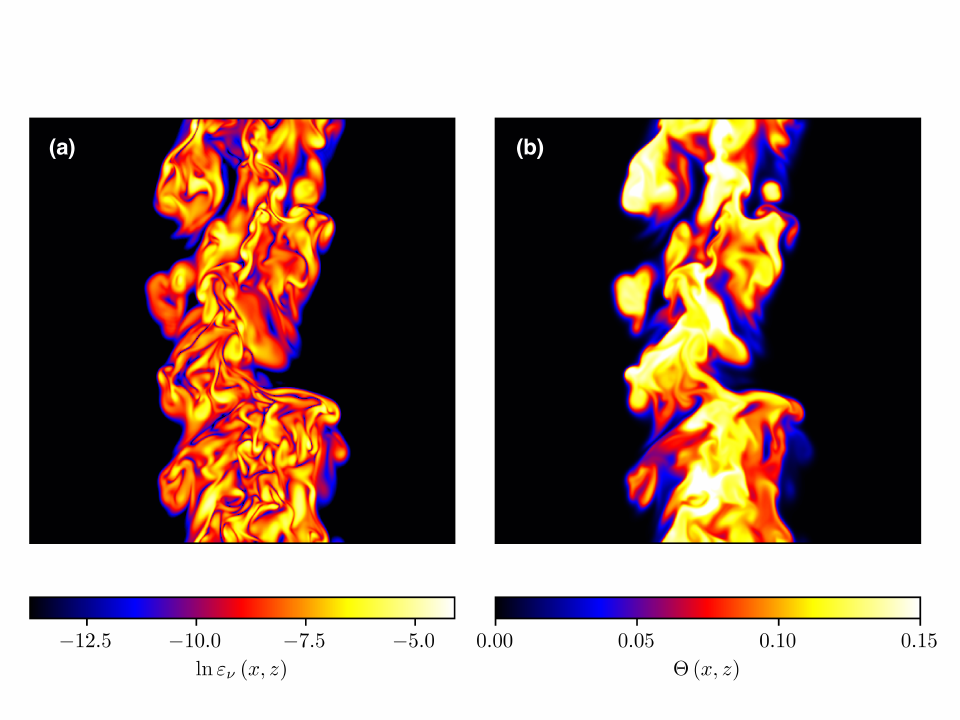}
\caption{Vertical sections in the $x$-$z$ plane of the 
dissipation field $\varepsilon_{\nu}$ (left) and of the
temperature field (right) from simulation D at time 
\textbf{$t=25$}.}
\label{fig1}
\end{figure}

\begin{figure}
\centering
\includegraphics[width=0.6\linewidth]{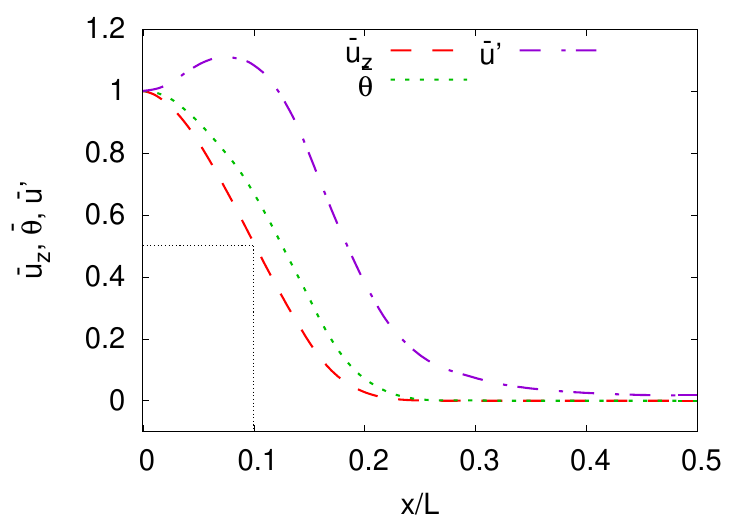}
\caption{Profiles of mean velocity $\overline{u_z}$ (dashed line), 
temperature $\overline{\theta}$ (dotted line) 
and velocity fluctuation $\overline{u'}$ (dash-dotted line) for the run $D$ at time 
$t=20$. Data are normalized using their centerline value.
The black dotted line represents the extension of $H_U$.}
\label{fig2}
\end{figure}

Before addressing the main results of the present work,
let us introduce the main flow features by considering the flow case D.
The instantaneous pattern taken by the flow is shown in figure~\ref{fig1}
through a vertical section of the instantaneous dissipation and temperature field.
A strongly convoluted shape of the plume boundaries is observed
that leads to entrainment of laminar flow regions up to the very core of the jet.
From a statistical point of view, the flow is characterized
by profiles of vertical velocity $\overline{u_z}(x,t)$,
temperature $\overline{\theta}(x,t)$ and velocity fluctuations
$\overline{u'}(x,t) = (\frac{1}{3}\overline{|{\bm u}-\overline{\bm u}|^2})^{1/2}$
that are shown in figure~\ref{fig2}.
A well-known bell shape is attained by the mean velocity and temperature field
with turbulent fluctuations that extend significantly
beyond the edge of the plume defined by mean profile themselves.

We also note that the width of the temperature profile is
slightly larger  than that of the vertical velocity. This phenomena is in
accordance with the typical value of the  turbulent Prandtl number,  which is
below 1 for jets and plumes \citep{craske2017turbulent}.

\section{Results}
\label{sec3}

\subsection{Time evolution of characteristic quantities}
\label{sec3.1}

\begin{figure}
\centering
\includegraphics[width=0.495\linewidth]{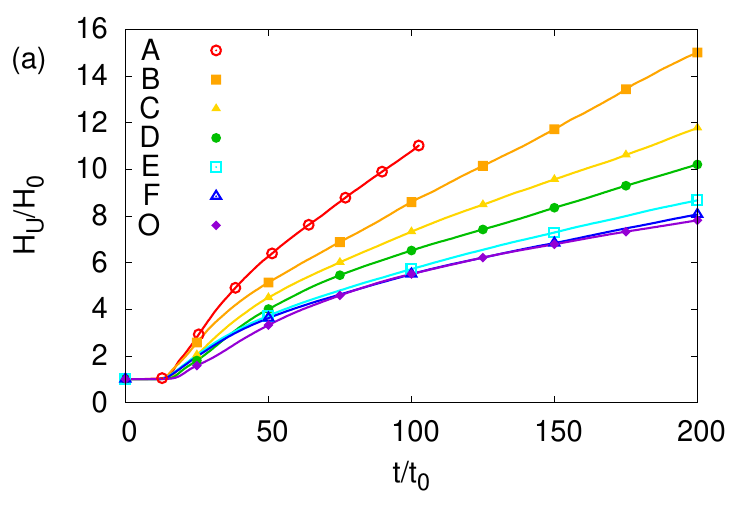}
\includegraphics[width=0.495\linewidth]{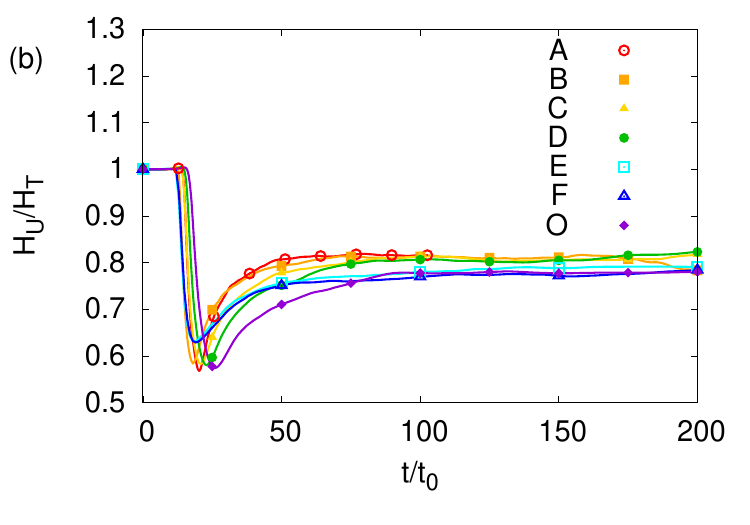}
\caption{(a) Temporal evolution of the forced plume width $H_U$ and
(b) temporal evolution of the ratio between the velocity and temperature 
widths $H_U$ and $H_T$. The time is rescaled with $t_0$.}
\label{fig3}
\end{figure}

At the initial times the perturbed laminar jet displays the onset of the Kelvin-Helmholtz
(KH) instability which eventually develops to a fully turbulent flow.
Before the instability the evolution of the jet is characterized by a 
constant width $H_U$, as shown in figure~\ref{fig3}a. 
The duration $t_D$ of this initial diffusive stage (defined by the 
region of constant $H_U$) appears to be 
dependent on the initial Reynolds number $Re_0$ and we 
empirically find that $t_{D} \simeq 15 t_0$ (note that diffusivity
is constant in our simulations).

Figure~\ref{fig3}b shows that in the turbulent regime the 
ratio $H_U/H_T$ is approximatively constant (around $0.8$).
For this reason, and for simplicity of the presentation, in the 
following we will use $H_U$ as the measure of the length scale.
Still, it can be observed that this ratio shows a weak dependence on the 
$Fr_0$, i.e. it is inversely proportional to it.

\begin{figure}
\centering
\includegraphics[width=0.6\linewidth]{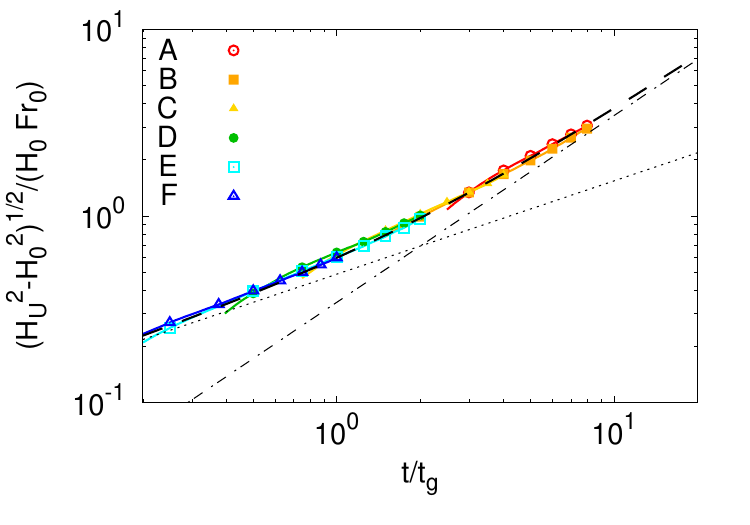}
\caption{
  Temporal evolution of the forced plume width
  rescaled according to the theoretical model
  (\ref{eq:plume_width_scaling}).
  The black dashed line represents the expression
  (\ref{eq:plume_width_scaling}) with $\alpha=0.237$.
We also show the asymptotic laws
in the momentum-dominated regime (\ref{eq:jetscaling})
(black dotted line)
and in the buoyancy-dominated regime (\ref{eq:plumescaling})
(black dash-dotted line).
The intersection of the two asymptotic laws occurs at $t/t_g=2$.}
\label{fig4}
\end{figure}

From figure~\ref{fig3}a it is difficult to distinguish between the initial 
shear regime (\ref{eq:jetscaling}) and the buoyancy regime 
(\ref{eq:plumescaling})
although it is evident that the growth of the mixing layer for
the three runs at the same $Re_0$ (A, B and C) is ordered as the
inverse of $Fr_0$ as predicted by (\ref{eq:plume_width_scaling_t0}).
Moreover, it is clear that the run O without buoyancy
does not display a transition to the linear growth, typical of the 
plume regime. 
Nonetheless, the model (\ref{eq:plume_width_scaling}) 
reproduces well the evolution of the mixing layer for all the cases considered. 
This is shown in figure~\ref{fig4} where we plot 
$(H_U^2(t)-H_0^2)/(H_0^2 Fr_0^2)$ as a function of the rescaled time $t/t_g$ 
together with the model (\ref{eq:plume_width_scaling}).
The value of the entrainment coefficient $\alpha=0.237$
has been obtained by fitting with a single curve the ensemble of data of all the simulations from A to F.
The initial times (before the development of the turbulent flow) have been excluded from the fit.
We remark that $\alpha$ is the only free parameter,
as the other quantities are defined \emph{a priori} and not measured.

Figure~\ref{fig4} shows that the transition between the two asymptotic regimes
  occurs over a broad range of time scales.
  Therefore, the cases considered in our study can be considered {\it transitional} cases.  
  In order to fully recover the asymptotic temporal scaling
  $(H_U(t)^2-H_0^2)^{1/2}/H_0Fr_0 = \sqrt{\alpha} (t/t_g)^{1/2}$ (dotted line in Fig.~\ref{fig4})
  of the jet regime (\ref{eq:jetscaling}) and
  $(H_U(t)^2-H_0^2)^{1/2}/H_oFr_0 = \sqrt{\alpha/2} (t/t_g)$ (dash-dotted line in Fig.~\ref{fig4})
  plume (\ref{eq:plumescaling}) regimes 
  it would be necessary to consider cases with larger and smaller $Fr_0$, respectively.  

\begin{figure}
\centering
\includegraphics[width=0.495\linewidth]{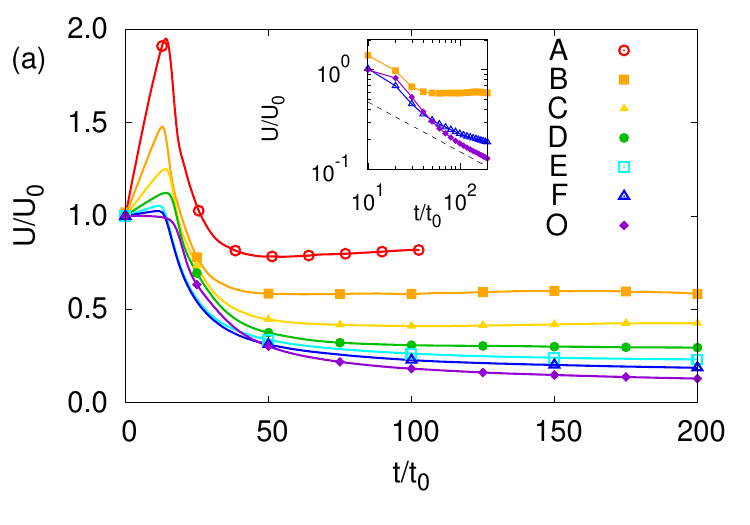}
\includegraphics[width=0.495\linewidth]{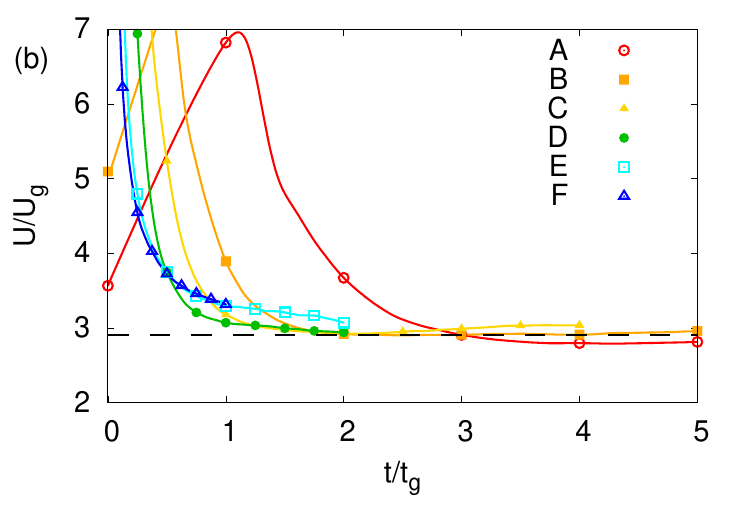}
\caption{Characteristic streamwise velocity amplitude $U$ as a function of $t/t_0$ (a).
The inset shows cases C, F and O on log-log axis, with a
dashed line showing the power-law $ U \sim t^{-1/2}$.
(b) Scaled values of $U$ by free-fall velocity $U_g$ as a function of $t/t_g$. 
Dashed line represent the asymptotic prediction $U/U_g = \sqrt{2/\alpha}$.
}
\label{fig5}
\end{figure}

The characteristic velocity $U(t)$  displays 
more clearly the transition from the jet regime (where 
$U \propto t^{-1/2}$) to the plume regime 
(with $U \sim U_g$). 
This is shown in figure~\ref{fig5}a where we can observe that, after 
the initial phase corresponding to the destabilization of the initial
laminar profile, the velocity amplitude reaches a constant
value for all the simulations with $U_g>0$.
On the contrary, the velocity amplitude for run O (i.e. in the 
absence of buoyancy forces) does not reach a finite asymptotic value.
These two different behaviors are even more clearly discernible in the inset of 
figure~\ref{fig5}a where the same quantity is shown in logarithmic scale
together with the jet scaling in equation (\ref{eq:jetscaling}). 
  The case O approaches asymptotically the jet scaling $U(t) \simeq t^{-1/2}$
  while the case $B$ clearly shows the plateau with constant value of $U$. 
  The case $F$ (with large but finite $Fr_0$) is initially similar to O
  but at long times it begins to display a convergence toward a finite value.
Velocity amplitudes rescaled with the theoretical predictions 
(\ref{eq:plumescaling}) are shown in figure~\ref{fig5}b.
In all cases, the characteristic velocity rescaled with $U_g$ 
  approaches at $t \gg t_g$ a constant value which is in quantitative agreement with the
  prediction $U/U_g = \sqrt{2/\alpha} \simeq 2.9$ discussed in Sect.\ref{sec2}. 
  We remind that the value of $\alpha$ (which fixes the asymptotic value of $U/U_g$) 
  depends on the specific definition of the amplitude $H_U(t)$.
We note also that the rescaling of times $t/t_g$
is not perfect because of the presence of the initial diffusive regime
which causes an initial offset $t_D$, which is not proportional to $t_g$.
A better collapse of the curves can be obtained by subtracting $t_D$.   
 
The deviation observed at short times for the case A deserves some comments.
In this case, due to the low initial values of
$Re_0$ and $Fr_0$, the flow accelerates much more than the other cases
before developing a turbulent behaviour. This results in a maximum velocity 
which is approximately twice as large as its initial value as 
shown in figure~\ref{fig5}a and in a longer time for reaching the 
asymptotic value. Indeed, better collapse can be obtained by defining
the buoyancy time $t_g$ with this maximum velocity instead than
$U_0$. 

\begin{figure}
\centering
\includegraphics[width=0.495\linewidth]{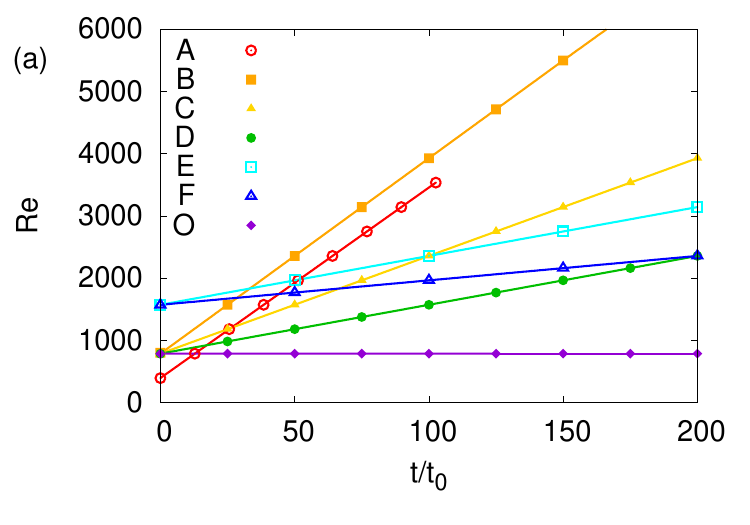}
\includegraphics[width=0.495\linewidth]{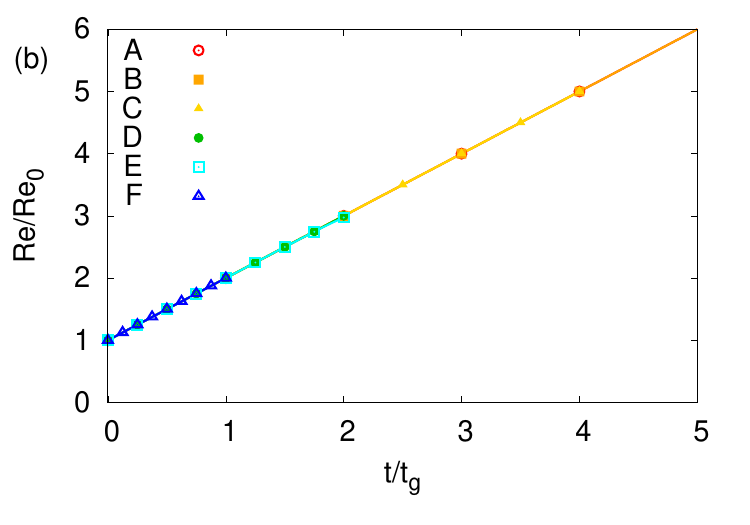}
\caption{Reynolds number $Re (t) = U H_U/\nu $ as a function of
  time $t/t0$. (a) Unscaled values.
  (b) Scaled values $Re/Re_0$ as a function of $t/t_g$.}
\label{fig6}
\end{figure}

The evolution of the Reynolds number of the flow, defined as 
$Re(t)=U H_U/\nu$ is shown in figure~\ref{fig6}. We see that in 
all runs with finite $Fr_0$ it increases with time, while 
for run O it remains constant.
  As discussed in Sect.\ref{sec2},
  this is a simple consequence of the fact that $Re(t)=Q(t)/\nu$
  and of (\ref{eq2.11}).
  By rescaling the time with $t_g$ and
  the values of $Re(t)$ with their initial value $Re_0$,
  the exact relation (\ref{eq_Ret}) is very well verified
  in our simulations as shown in figure~\ref{fig6}b.

\begin{figure}
\centering
\includegraphics[width=0.495\linewidth]{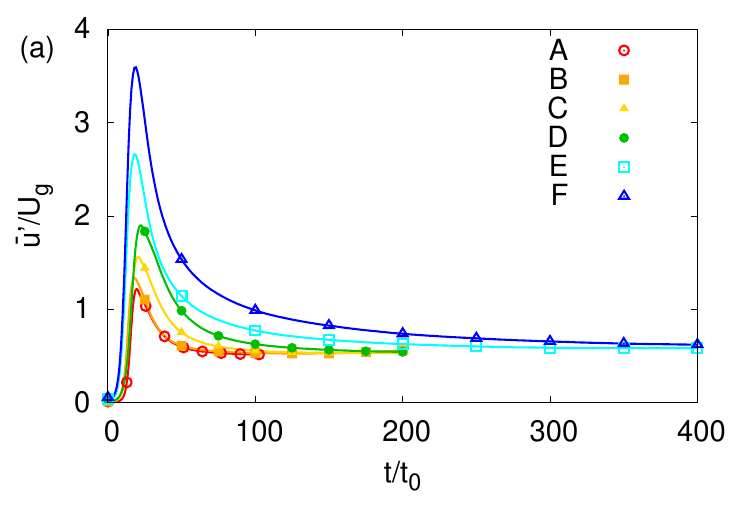}
\includegraphics[width=0.495\linewidth]{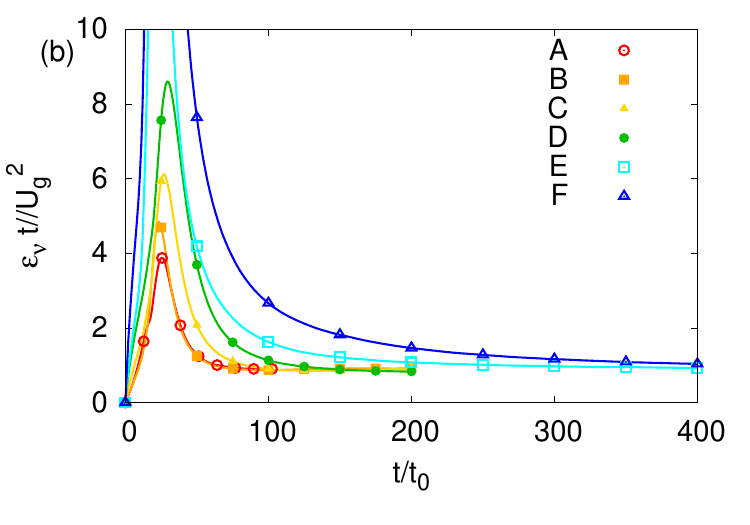}
\caption{(a) Amplitude of velocity fluctuations $u'$ rescaled with $U_g$ 
as a function of $t/t_0$.
(b) Kinetic energy dissipation rate $\varepsilon_\nu$ compensated with 
the prediction $U_g^2/t$ (\ref{eq:plumescaling_turb}) 
as a function of $t/t_0$. 
}
\label{fig7}
\end{figure}

\begin{figure}
\centering
\includegraphics[width=0.495\linewidth]{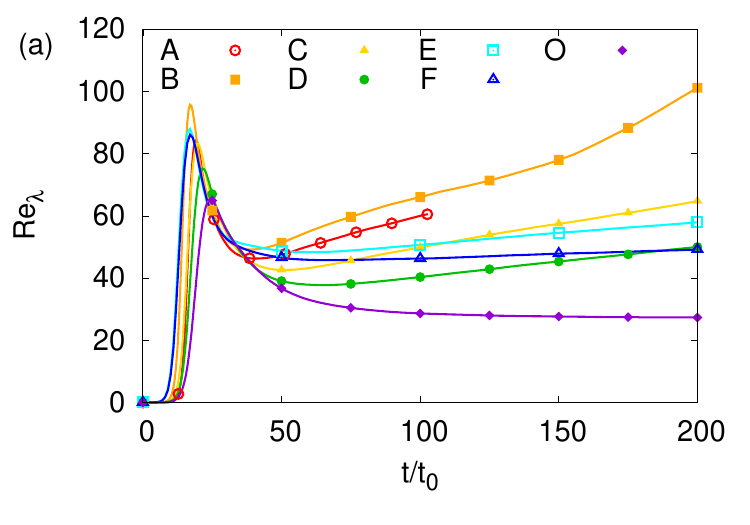}
\includegraphics[width=0.495\linewidth]{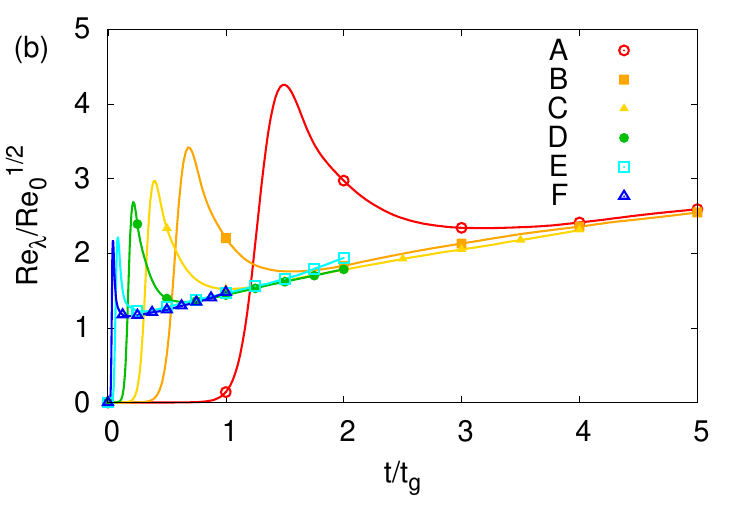}
\caption{Taylor-microscale Reynolds number as a function of $t/t0$ (a) Unscaled values.
  (b) Scaled values by the initial Reynolds number $Re_0$, as a function of $t/t_g$.}
\label{fig8}
\end{figure}

Finally, we consider the time evolution of turbulent quantities.
As discussed in Sect.~\ref{sec2}, the RMS turbulent velocity fluctuations
$u' = \langle \frac{1}{3} |{\bm u}-\overline{\bm u}|^2 \rangle^{1/2}$,
are expected to scale in time as the characteristic velocity $U(t)$
and therefore to reach an asymptotic constant value in the plume regime.
Of course, at variance with the characteristic velocity, fluctuations are
zero during the initial laminar phase.
  As shown in figure~\ref{fig7}a,
  the convergence toward an asymptotic value is observed for all the runs. 
  At long times, the relative intensity of turbulent fluctuations 
  $u'/U$ approaches a constant value $u'/U \simeq 0.18$. 

Average energy dissipation
$\varepsilon_{\nu} = \nu \langle \left( \partial_i u_j \right)^2 \rangle$
is plotted in figure~\ref{fig7}b, compensated with the dimensional scaling
{$\varepsilon_{\nu} \propto U_g^2 t^{-1}$ expected for the buoyancy dominated
regime (Eq.~\ref{eq:plumescaling_turb}).  
As for velocity fluctuations, also in this case we observe
the presence of a peak in correspondence of the development of turbulence. 
When comparing with figure~\ref{fig3}b we see that this peak appears at
the time at which the evolution of the flow enters in the 
self-similar regime.

  From the values of $u'$ and $\varepsilon_\nu$ it is straightforward to obtain the
  evolution of the Taylor-microscale Reynolds number
  $Re_\lambda =u'^2\sqrt{15/ \nu \varepsilon_\nu}$, which is shown in Fig.~\ref{fig8}a.
  After the initial diffusive stage, in which $Re_\lambda$ is zero  
  we observe a rapid increase of the value of $Re_\lambda$ up to a peak value 
  which is achieved at almost the same time of the maximum of $u'$ observed in Fig.~\ref{fig7}a.  
  In the simulation with $Fr_0 = \infty$ (case O) $Re_\lambda$ attains asymptotically a constant value, 
  in agreement with the jet-scaling prediction (\ref{eq:jetscaling_turb}).
  A plateau corresponding to the jet scaling regime is also observed at intermediate times
  $50 t_0 \lesssim t \lesssim 150 t_0$ in the simulation with the largest (finite) $Fr_0$
  (case F).
  In all the other cases (A-E), $Re_\lambda$ grows in time the turbulent regime.  
  The convergence toward the asymptotic buoyancy-dominated regime is shown in Fig.~\ref{fig7}b,
  in which the rescaled values  $Re_\lambda/Re_0^{1/2}$ are plotted as a function of $t/t_g$,
  according to (\ref{eq:plumescaling_turb}).  

\subsection{Velocity and temperature profiles}
\label{sec3.2}

We now consider the mean velocity and temperature profiles. 
Since the total thermal content $C(t) = T(t)H(t)$ is a conserved
quantity in both the shear-dominated and buoyancy-dominated regimes,
we expect that the mean temperature profiles $\overline{\theta}(x,t)$
at different times can be simply rescaled by
$H_T(t)$ and the product $ T_0 H_0$. Figure~\ref{fig9} confirms this
hypothesis, indicating a remarkable self-similar behaviour of the temperature 
profiles, with almost no differences between the two regimes.

On the basis of the hypothesis of self-similarity of the flow,
we expect that also the profiles of the vertical component of the velocity $\overline{u_z}(x,t)$,
can be collapsed at all time by rescaling them with the scale 
$H_U(t)$ and characteristic velocity $U(t)$ obtained from the model 
(\ref{eq:plume_width_scaling}) and (\ref{eq:plume_velocity_scaling})
discussed in Sect.\ref{sec2}.
In Fig.~\ref{fig10} we show the rescaled profiles at different times in the
shear-dominated regime for the set D of simulations (panel a)
and in the buoyancy-dominated regime for the set B of simulations (panel b).
In almost all cases we observe a good collapse to a self-similar profile.
Note that here we used the same value of the parameter $\alpha =0.237$
which has been previously determined by fitting the time evolution of $H_U(t)$,
therefore the collapse of the profiles is achieved without further fitting parameters. 
The small deviations which are observed at short times $t= 0.5t_g$ in the case D
can be ascribed to the fact that the turbulent regime is not yet fully achieved at that time.
This is consistent with Fig.~\ref{fig9}b which shows that in the case D the asymptotic regime of 
$Re_\lambda$ is achieved at times $t > 0.5t_g$.
It is interesting to notice that in the buoyancy-dominated regime, 
analogously to the case of the passive scalar field advected by the jet, the self-similar shapes 
of the temperature and velocity profiles are different from each other.

\begin{figure}
\centering
\includegraphics[width=0.495\linewidth]{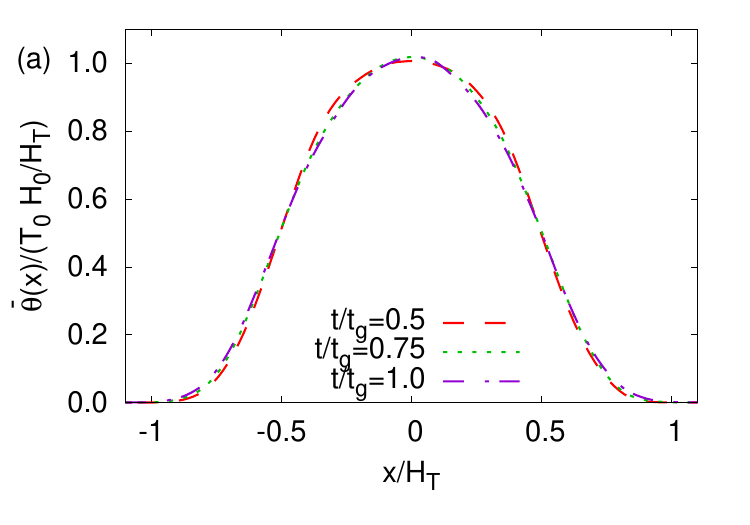}
\includegraphics[width=0.495\linewidth]{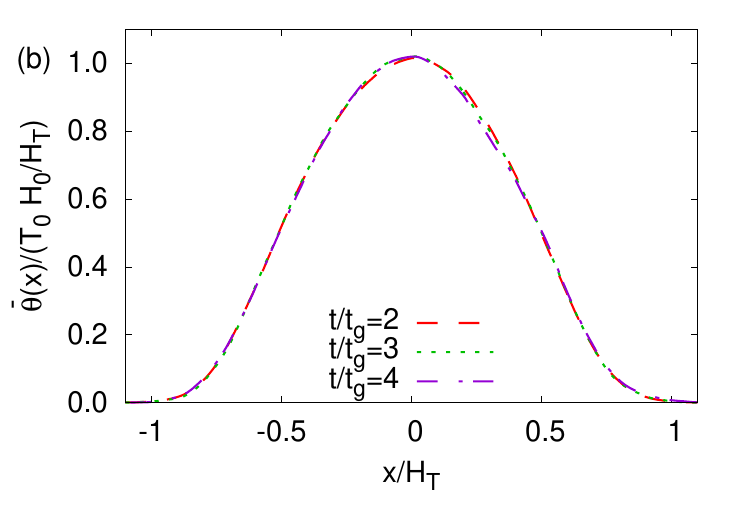}
\caption{Mean temperature profiles $\overline{\theta}(x,t)$ for the set C 
of simulations,
at different times, in the (a) shear- and (b) buoyancy-dominated regime, rescaled by the initial total thermal content $C_0 = T_0 H_0$ and the characteristic width $H_T(t)$.}
\label{fig9}
\end{figure}

\begin{figure}
\centering
\includegraphics[width=0.495\linewidth]{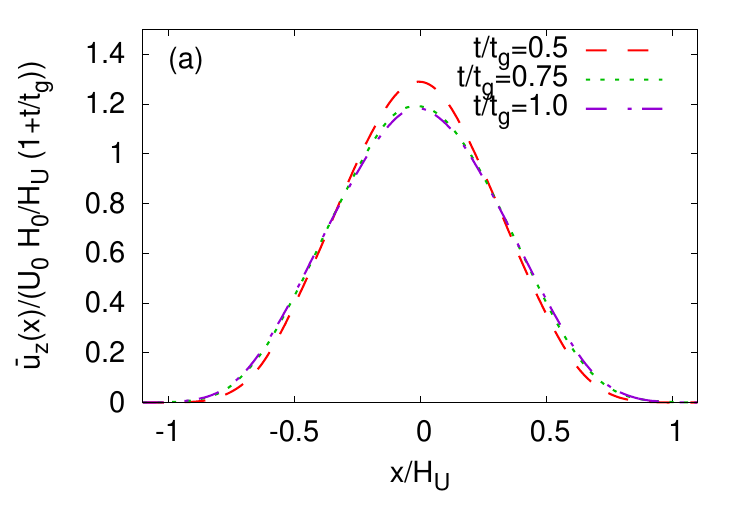}
\includegraphics[width=0.495\linewidth]{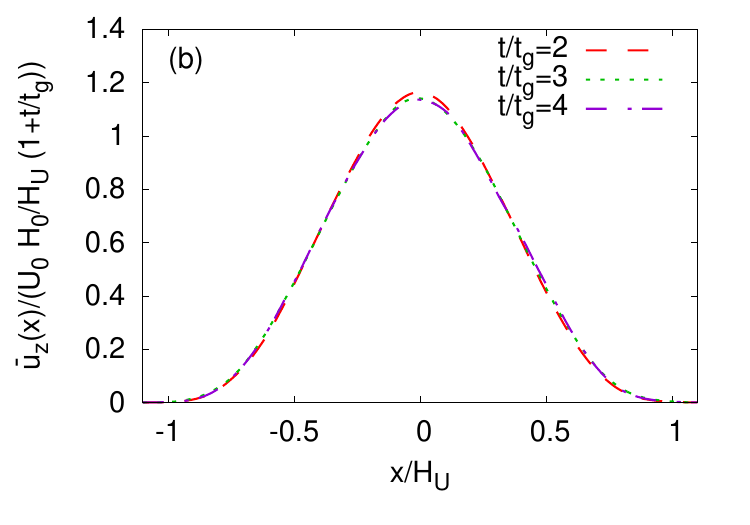}
\caption{Mean profiles of vertical velocity $\overline{u_z}(x,t)$ for the set 
D of simulations, at different times, in the (a) shear--dominated regime (a)
and for the set B of simulations in the buoyancy-dominated regime (b),
rescaled by the characteristic width $H_U$ and velocity $U(t)$
  obtained by the model (\ref{eq:plume_width_scaling})
  and (\ref{eq:plume_velocity_scaling}) with $\alpha=0.237$.}
\label{fig10}
\end{figure}

\section{Conclusions}
\label{sec4}
The transition from momentum-dominated to buoyancy-dominated regime in buoyant
forced plumes is studied in the present work by means of direct numerical
simulation. 
By considering the time evolution of the total thermal content and of the volume flux, 
we obtain a general model for the temporal growth of 
the width of the plume $H_U(t)$ (\ref{eq:plume_width_scaling}) 
and for its characteristic velocity $U(t)$ (\ref{eq:plume_velocity_scaling})  
which allows to identify two different scaling regimes, separated by a crossover time $t_g$. 
In the limit of large $Re$ and $Fr$ numbers, the model (\ref{eq:plume_width_scaling}, \ref{eq:plume_velocity_scaling}) 
prescribes that for $t\ll t_g$ the plume dynamic should behave as a 
non-buoyant jet with scaling laws $H_U(t)\propto t^{1/2}$ and $U(t) \propto t^{-1/2}$ 
while at later times $t \gg t_g$ the system retrieves the plume-like 
behaviour with $H_u(t)\propto t$ and $U(t)\simeq constant$.  
A strong point of our model is that it provides a simple, yet effective parametrisation of the transition between the two regimes, 
with a single fitting parameter, represented by the entrainment coefficient. 

The results of the numerical simulations shows that our model is indeed able to capture correctly the transitional regime 
which is observed at finite $Fr$ and $Re$,  
This is confirmed also by the time evolution of the velocity profiles, 
which can be collapsed using the characteristic scale and velocity provided by the model. 
We also found that the asymptotic buoyant temporal scaling laws are always retrieved at 
late times while the jet-like scaling laws are not clearly observed in the simulations at finite Froude number.  
Indeed, buoyancy plays a crucial role in the destabilization and the growth of
the plume width, also influencing its initial stage, hence causing deviations
from the pure jet behaviour. 
The jet behaviour is clearly observed in the non-buoyant case, thus suggesting
that, in general, one needs a very large $Fr_0$ (i.e. large transition time
$t_g$) in order to disentangle the two scaling laws in (\ref{eq:plume_width_scaling}). 
By increasing the initial $Re_0$ while keeping a constant $Fr_0$ (e.g. decreasing $\nu$), 
it is possible to reduce the initial diffusive stage duration $t_D$, anticipating the transition to the turbulent regime. Nevertheless, in order to observe a clear jet scaling, $t_g/t_0 \gg 1$ is required, implying that a high $Re_0$ alone is not sufficient to induce a pure jet behaviour in the initial stage.
This suggests a possible line of research for future works.

\section*{Acknowledgments}

We acknowledge {\it HPC CINECA} for computing resources 
(INFN-CINECA Grant No. INFN23-FieldTurb). L.P. 
thanks {\it Fondazione CRT} for financial support 
within the project 2021.0858.
M.C.-E. acknowledges the financial support given by the Department of
Engineering ``Enzo Ferrari'' of the University of Modena and Reggio 
Emilia through the action {\it FAR dipartimentale 2024/2025}.
We also thank one of the Referees for very useful remarks which helped to
improve the manuscript.

The authors report no conflict of interest.

\label{sec4}
\bibliographystyle{jfm}
\bibliography{bibfile.bib}

\begin{thebibliography}{32}
\expandafter\ifx\csname natexlab\endcsname\relax\def\natexlab#1{#1}\fi
\def\au#1{#1} \def\ed#1{#1} \def\yr#1{#1}\def\at#1{#1}\def\jt#1{\textit{#1}}
  \def\bt#1{#1}\def\bvol#1{\textbf{#1}} \def\vol#1{#1} \def\pg#1{#1}
  \def\publ#1{#1}\def\arxiv#1{#1}\def\org#1{#1}\def\st#1{\textit{#1}}

\bibitem[Akula {\em et~al.\/}(2017)Akula, Suchandra, Mikhaeil \&
  Ranjan]{akula2017dynamics}
{\sc \au{Akula, Bhanesh}, \au{Suchandra, Prasoon}, \au{Mikhaeil, Mark} \&
  \au{Ranjan, Devesh}} \yr{2017}  \at{{Dynamics of unstably stratified free
  shear flows: an experimental investigation of coupled Kelvin--Helmholtz and
  Rayleigh--Taylor instability}}.  \jt{Journal of Fluid Mechanics}  \bvol{816},
   \pg{619--660}.

\bibitem[Bisset {\em et~al.\/}(2002)Bisset, Hunt \& Rogers]{Bisset_2002}
{\sc \au{Bisset, David~K.}, \au{Hunt, Julian C.~R.} \& \au{Rogers, Michael~M.}}
  \yr{2002}  \at{{The turbulent/non-turbulent interface bounding a far wake}}.
  \jt{Journal of Fluid Mechanics}  \bvol{451},  \pg{383–410}.

\bibitem[Borrell \& Jiménez(2016)]{Borrell_Jimenez_2016}
{\sc \au{Borrell, Guillem} \& \au{Jiménez, Javier}} \yr{2016}  \at{{Properties
  of the turbulent/non-turbulent interface in boundary layers}}.  \jt{Journal
  of Fluid Mechanics}  \bvol{801},  \pg{554–596}.

\bibitem[Van~den Bremer \& Hunt(2014)]{van2014two1}
{\sc \au{Van~den Bremer, TS} \& \au{Hunt, Gary~R}} \yr{2014}
  \at{{Two-dimensional planar plumes and fountains}}.  \jt{Journal of fluid
  mechanics}  \bvol{750},  \pg{210--244}.

\bibitem[Brizzolara {\em et~al.\/}(2023)Brizzolara, Mollicone, van Reeuwijk \&
  Holzner]{brizzolara2023entrainment}
{\sc \au{Brizzolara, Stefano}, \au{Mollicone, Jean-Paul}, \au{van Reeuwijk,
  Maarten} \& \au{Holzner, Markus}} \yr{2023}  \at{{Entrainment at multi-scales
  in shear-dominated and Rayleigh--Taylor turbulence}}.  \jt{European Journal
  of Mechanics-B/Fluids}  \bvol{101},  \pg{294--302}.

\bibitem[Brizzolara {\em et~al.\/}(2021)Brizzolara, Mollicone, van Reeuwijk,
  Mazzino \& Holzner]{brizzolara2021transition}
{\sc \au{Brizzolara, Stefano}, \au{Mollicone, Jean-Paul}, \au{van Reeuwijk,
  Maarten}, \au{Mazzino, Andrea} \& \au{Holzner, Markus}} \yr{2021}
  \at{{Transition from shear-dominated to Rayleigh--Taylor turbulence}}.
  \jt{Journal of Fluid Mechanics}  \bvol{924},  \pg{A10}.

\bibitem[Cimarelli \& Boga(2021)]{cimarelli2021numerical}
{\sc \au{Cimarelli, A} \& \au{Boga, G}} \yr{2021}  \at{{Numerical experiments
  on turbulent entrainment and mixing of scalars}}.  \jt{J. Fluid Mech.}
  \bvol{927},  \pg{A34}.

\bibitem[Cimarelli {\em et~al.\/}(2021)Cimarelli, Mollicone, Van~Reeuwijk \&
  De~Angelis]{cimarelli2021spatially}
{\sc \au{Cimarelli, A}, \au{Mollicone, J-P}, \au{Van~Reeuwijk, M} \&
  \au{De~Angelis, E}} \yr{2021}  \at{{Spatially evolving cascades in temporal
  planar jets}}.  \jt{J. Fluid Mech.}  \bvol{910},  \pg{A19}.

\bibitem[Corrsin \& Kistler(1955)]{Corrsin_TNTI}
{\sc \au{Corrsin, Stanley} \& \au{Kistler, Alan~L.}} \yr{1955}
  \at{{Free-stream boundaries of turbulent flows}}.  \jt{NACA Tech. Rep.}
  \bvol{TN-1244}.

\bibitem[Craske {\em et~al.\/}(2017)Craske, Salizzoni \&
  Van~Reeuwijk]{craske2017turbulent}
{\sc \au{Craske, John}, \au{Salizzoni, Pietro} \& \au{Van~Reeuwijk, Maarten}}
  \yr{2017}  \at{{The turbulent Prandtl number in a pure plume is 3/5}}.
  \jt{Journal of Fluid Mechanics}  \bvol{822},  \pg{774--790}.

\bibitem[De~Rooy {\em et~al.\/}(2013)De~Rooy, Bechtold, Fr{\"o}hlich,
  Hohenegger, Jonker, Mironov, Pier~Siebesma, Teixeira \&
  Yano]{de2013entrainment}
{\sc \au{De~Rooy, Wim~C}, \au{Bechtold, Peter}, \au{Fr{\"o}hlich, Kristina},
  \au{Hohenegger, Cathy}, \au{Jonker, Harm}, \au{Mironov, Dmitrii},
  \au{Pier~Siebesma, A}, \au{Teixeira, Joao} \& \au{Yano, Jun-Ichi}} \yr{2013}
  \at{{Entrainment and detrainment in cumulus convection: An overview}}.
  \jt{Quarterly Journal of the Royal Meteorological Society}  \bvol{139}~(670),
   \pg{1--19}.

\bibitem[Hunt \& Van~den Bremer(2011)]{hunt2011classical}
{\sc \au{Hunt, GR} \& \au{Van~den Bremer, TS}} \yr{2011}  \at{{Classical plume
  theory: 1937--2010 and beyond}}.  \jt{IMA journal of applied mathematics}
  \bvol{76}~(3),  \pg{424--448}.

\bibitem[Krug {\em et~al.\/}(2017)Krug, Chung, Philip \&
  Marusic]{krug2017global}
{\sc \au{Krug, Dominik}, \au{Chung, Daniel}, \au{Philip, Jimmy} \& \au{Marusic,
  Ivan}} \yr{2017}  \at{{Global and local aspects of entrainment in temporal
  plumes}}.  \jt{Journal of Fluid Mechanics}  \bvol{812},  \pg{222--250}.

\bibitem[Mathew \& Basu(2002)]{Mathew2002}
{\sc \au{Mathew, Joseph} \& \au{Basu, Amit~J.}} \yr{2002}  \at{{Some
  characteristics of entrainment at a cylindrical turbulence boundary}}.
  \jt{Physics of Fluids}  \bvol{14}~(7),  \pg{2065--2072}.

\bibitem[Mazzino \& Rosti(2021)]{mazzino2021unraveling}
{\sc \au{Mazzino, Andrea} \& \au{Rosti, Marco~Edoardo}} \yr{2021}
  \at{{Unraveling the secrets of turbulence in a fluid puff}}.  \jt{Physical
  Review Letters}  \bvol{127}~(9),  \pg{094501}.

\bibitem[Morton(1959)]{morton1959forced}
{\sc \au{Morton, BR103670}} \yr{1959}  \at{{Forced plumes}}.  \jt{Journal of
  Fluid mechanics}  \bvol{5}~(1),  \pg{151--163}.

\bibitem[Morton \& Middleton(1973)]{morton1973scale}
{\sc \au{Morton, BR} \& \au{Middleton, Jason}} \yr{1973}  \at{{Scale diagrams
  for forced plumes}}.  \jt{Journal of Fluid Mechanics}  \bvol{58}~(1),
  \pg{165--176}.

\bibitem[Paillat \& Kaminski(2014)]{paillat2014entrainment}
{\sc \au{Paillat, S} \& \au{Kaminski, E}} \yr{2014}  \at{{Entrainment in plane
  turbulent pure plumes}}.  \jt{Journal of Fluid Mechanics}  \bvol{755},
  \pg{R2}.

\bibitem[van Reeuwijk \& Holzner(2014)]{vanReeuwijk_Holzner_2014}
{\sc \au{van Reeuwijk, Maarten} \& \au{Holzner, Markus}} \yr{2014}  \at{{The
  turbulence boundary of a temporal jet}}.  \jt{Journal of Fluid Mechanics}
  \bvol{739},  \pg{254–275}.

\bibitem[da~Silva {\em et~al.\/}(2014)da~Silva, Hunt, Eames \&
  Westerweel]{daSilvaAnnuRev2014}
{\sc \au{da~Silva, Carlos~B.}, \au{Hunt, Julian~C.R.}, \au{Eames, Ian} \&
  \au{Westerweel, Jerry}} \yr{2014}  \at{{Interfacial Layers Between Regions of
  Different Turbulence Intensity}}.  \jt{Annual Review of Fluid Mechanics}
  \bvol{46}~(1),  \pg{567--590}.

\bibitem[da~Silva \& Pereira(2008)]{da2008invariants}
{\sc \au{da~Silva, Carlos~B} \& \au{Pereira, Jos{\'e}~CF}} \yr{2008}
  \at{{Invariants of the velocity-gradient, rate-of-strain, and
  rate-of-rotation tensors across the turbulent/nonturbulent interface in
  jets}}.  \jt{Physics of fluids}  \bvol{20}~(5).

\bibitem[da~Silva \& Taveira(2010)]{daSilva_2010}
{\sc \au{da~Silva, Carlos~B.} \& \au{Taveira, Rodrigo~R.}} \yr{2010}  \at{{The
  thickness of the turbulent/nonturbulent interface is equal to the radius of
  the large vorticity structures near the edge of the shear layer}}.  \jt{Phys.
  Fluids}  \bvol{22},  \pg{121702}.

\bibitem[Snider \& Andrews(1994)]{snider1994rayleigh}
{\sc \au{Snider, Dale~M} \& \au{Andrews, Malcolm~J}} \yr{1994}
  \at{{Rayleigh--Taylor and shear driven mixing with an unstable thermal
  stratification}}.  \jt{Physics of Fluids}  \bvol{6}~(10),  \pg{3324--3334}.

\bibitem[Speer \& Rona(1989)]{speer1989model}
{\sc \au{Speer, Kevin~G} \& \au{Rona, Peter~A}} \yr{1989}  \at{{A model of an
  Atlantic and Pacific hydrothermal plume}}.  \jt{Journal of Geophysical
  Research: Oceans}  \bvol{94}~(C5),  \pg{6213--6220}.

\bibitem[Van Den~Bremer \& Hunt(2014)]{van2014two}
{\sc \au{Van Den~Bremer, TS} \& \au{Hunt, GR}} \yr{2014}  \at{{Two-dimensional
  planar plumes: non-Boussinesq effects}}.  \jt{Journal of fluid mechanics}
  \bvol{750},  \pg{245--258}.

\bibitem[Van~Reeuwijk \& Craske(2015)]{van2015energy}
{\sc \au{Van~Reeuwijk, Maarten} \& \au{Craske, John}} \yr{2015}
  \at{{Energy-consistent entrainment relations for jets and plumes}}.
  \jt{Journal of Fluid Mechanics}  \bvol{782},  \pg{333--355}.

\bibitem[Watanabe {\em et~al.\/}(2015)Watanabe, Sakai, Nagata, Ito \&
  Hayase]{Watanabe_shearlayer}
{\sc \au{Watanabe, T.}, \au{Sakai, Y.}, \au{Nagata, K.}, \au{Ito, Y.} \&
  \au{Hayase, T.}} \yr{2015}  \at{{Turbulent mixing of passive scalar near
  turbulent and non-turbulent interface in mixing layers}}.  \jt{Physics of
  Fluids}  \bvol{27}~(8),  \pg{085109}.

\bibitem[Watanabe {\em et~al.\/}(2018)Watanabe, Zhang \& Nagata]{Watanabe2018}
{\sc \au{Watanabe, T.}, \au{Zhang, X.} \& \au{Nagata, K.}} \yr{2018}
  \at{{Turbulent/non-turbulent interfaces detected in DNS of incompressible
  turbulent boundary layers}}.  \jt{Phys. Fluids}  \bvol{30},  \pg{035102}.

\bibitem[Westerweel {\em et~al.\/}(2005)Westerweel, Fukushima, Pedersen \&
  Hunt]{Westerweel_2005}
{\sc \au{Westerweel, J}, \au{Fukushima, C}, \au{Pedersen, JM} \& \au{Hunt,
  JCR}} \yr{2005}  \at{{Mechanics of the turbulent-nonturbulent interface of a
  jet}}.  \jt{Phys. Rev. Lett.}  \bvol{95},  \pg{174501}.

\bibitem[Westerweel {\em et~al.\/}(2009)Westerweel, Fukushima, Pedersen \&
  Hunt]{Westernweel2009}
{\sc \au{Westerweel, J.}, \au{Fukushima, C.}, \au{Pedersen, J.~M.} \& \au{Hunt,
  J. C.~R.}} \yr{2009}  \at{{Momentum and scalar transport at the
  turbulent/non-turbulent interface of a jet}}.  \jt{Journal of Fluid
  Mechanics}  \bvol{631},  \pg{199--230}.

\bibitem[Woods(2010)]{woods2010turbulent}
{\sc \au{Woods, Andrew~W}} \yr{2010}  \at{{Turbulent plumes in nature}}.
  \jt{Annual Review of Fluid Mechanics}  \bvol{42},  \pg{391--412}.

\bibitem[Zeldovich(1937)]{zeldovich1937asymptotic}
{\sc \au{Zeldovich, Ya~B}} \yr{1937}  \at{{The asymptotic laws of
  freely-ascending convective flows}}.  \jt{Zhur. Eksper. Teor. Fiz}
  \bvol{7}~(12),  \pg{1463--1465}.

\end{thebibliography}


\end{document}